# From naive to sophisticated behavior in multiagents based financial market models.

**R. Mansilla[1,2]**

**[1]Department of Complex Systems**
**Physical Institute**
**National University of Mexico**

**[2]Department of Differential Equations**
**Faculty of Mathematics and Computer Science**
**University of Havana, Cuba**

**Abstract**

We discuss the behavior of two magnitudes, physical complexity and mutual information function of the outcome of a model of heterogeneous, inductive rational agents inspired in the El Farol Bar problem and the Minority Game. The first is a measure rooted in Kolmogorov-Chaitin theory and the second one a measure related with Shannon's information entropy. We make extensive computer simulations , as result of which, we propose an *ansatz* for physical complexity of the type $C(l) = l^{\alpha}$ and establish the dependence of exponent $\alpha$ from the parameters of the model. We discuss the accuracy of our results and the relationship with the behavior of mutual information function as a measure of time correlations of agents choice.

## 1 Introduction.

In many natural and social systems agents establish among them self a complex network of interactions. Often in such systems it is the case that successful agents are those which act in ways that are distinct from their competitors. El Farol bar problem [1] has become a paradigm of such kind of models. The next step in this endeavor is the well known Minority Game model [2], [3]. Numerical simulations by several authors [2], [4-8] have shown that minority game (MG) displays a remarkably rich emergent collective behavior.

The setup of the minority game is the following: The $N$ agents have to choose at each time step whether to go in room -1 or 1. The agents who have chosen the less crowded room (minority room) win, and the other loose. The agents have limited capabilities, and only "remember" the last $m$ outcomes of the game. The number $m$ is called memory size or brain size. In order to decide in what room to go agents use strategies. It is the way inductive reasonning is made. A strategy is a choosing device, that is, an object that process the outcomes of the winning room in the last $m$ time step and accordingly to this information prescribes in what room to go the next one.

The agents randomly pick $s$ strategies at the beginning of the game. After each turn, the agents assign one (virtual) point to each of his strategies which would have predicted the correct outcome. At each turn of the game, they use whichever is the most successful strategy among the s in his possession, i.e., he choose the one that has gained most virtual points. Variations of this setup are studied in the remarkable paper [9].

In order to understand the complex behavior of the model, particular emphasis have been devoted to study the mean square deviation of the number of agents making a given choice $\sigma$. In the financial context, this observable is called volatility. Although the great amount of papers devoted to study this quantity (see [4] and [5] for example), A. Cavagna [6] proved that the behavior of the above mentioned measure does not depend of the real history of the game. Hence, we proposed [10] an alternative new approach for the study of the complex behavior of minority game borrowing tools from thermodynamics and statistical physics [10].

The above mentioned models capture some “stylized facts” of real financial markets, because due to formidable complexity of their behavior, deductive rationality is unrealistic in such strategic situations. It is generally accepted that agents follow more likely strategies which have been more successful in the past. The well know “law of efect” [11] reflect this kind of behavior.

In the original minority game agents make inductive reasoning but only reward each strategy in an imperfect way. They do not take into account the impact of their choice in the macroscopical observable. In that sense agents are *naive*. In the paper [9] another kind of behaviors are studied which yield more sophisticated conduct of agents. It depend on the way agents reward all their strategies and how they estimate their impact on the market index.

The aim of this paper is to study how physical complexity and mutual information function of the outcomes of the game change, when the conduct of agents vary from naive to more sophisticated behavior. A measure of this change $\varphi$ is defined and we establish an *ansatz* for the physical complexity $C(l) = l^{\alpha(m,s,\varphi)}$. We make extensive computer simulations to show that $\alpha(m,s,\varphi) = a(m,s)\varphi + b(m,s)$ and calculate approximated expression for $a(m,s)$ and $b(m,s)$. We also discuss the implication of that behavior of $C(l)$ in the form of the mutual information function.

The paper is structured as follows. In Sec. 2 we introduce our measure of the sophistication of the conduct of the agents and discuss the relationship with similar result obtained in [9]. We also breifly review the result defined in [10] and establish an *ansatz* for physical complexity. We discuss how the form of the exponent is related with the efficiency of the market and with the memory size and the number of strategies used by agents. In Sec. 3 we study the implication of the above result on the behavior of mutual information function of the string of outcomes of the model. Section 4 is for conclusions and in Sec. 5 one could find the references.

## 2 The Behavior of Physical Complexity.

We first breifly review some concepts. Physical complexity (first studied in [12-14]) is defined as the number of binary digits that are meaningful in a string $\eta$ with respect to some enviroment $\varepsilon$. As in [10] we take in the capacity of $\varepsilon$, the stream of data composed by the outcomes of the model and study the physical complexity of substrings $\eta$ of $\varepsilon$,. In terms of Kolmogorov-Chaitin entropy, physical complexity can be written as:

$$K(\eta : \varepsilon) = |\eta| - K(\eta / \varepsilon) \quad (1)$$

where $K(\eta / \varepsilon)$ [14] is the length of the smallest program that computes $\eta$ in a Turing machine having $\varepsilon$ as input:

$$K(\eta / \varepsilon) = min\left\{|\pi| : \eta = C_T(\pi, \varepsilon)\right\} \quad (2)$$

Notice that $|\eta|$ also represent (see [10] for details) the unconditional complexity of string $\eta$ i.e., the value of complexity if the input would be $\varepsilon = \varnothing$. Of course, the measure $K(\eta : \varepsilon)$ as defined in Eq. 3 has few practical application, mainly because it is impossible to know the way in which information about $\varepsilon$ is coded in $\eta$. However (as proved in [14]), if a statistical ensemble of strings is available to us, then the determination of complexity becomes an exercise in information theory. It can be proved that the average values $C(|\eta|)$ of the physical complexity $K(\eta:\varepsilon)$ taken over an ensemble $\Sigma$ of strings of length $|\eta|$ can be approximated by:

$$C(|\eta|) = \langle K(\eta:\varepsilon)\rangle_{\Sigma} \cong |\eta| - K(\Sigma/\varepsilon) \tag{3}$$

where:

$$K(\Sigma/\varepsilon) = -\sum_{\eta\in\Sigma} p(\eta/\varepsilon)\log_2 p(\eta/\varepsilon) \tag{4}$$

and the sum is taking over all the strings $\eta$ in the ensemble $\Sigma$. In a population of $N$ strings in the enviroment $\varepsilon$ , the quantity $\frac{n(\eta)}{N}$, where $n(s)$ denotes the number of strings equal to $\eta$ in $\Sigma$, approximates $p(\eta/\varepsilon)$ as $N \to \infty$.

Let $\varepsilon = a_1 a_2 a_3 \cdots a_n \cdots$; $a_i \in \{-1, 1\}$ be the stream of outcomes of the game and $l$ a positive integer $l \geq 2$. Let $\Sigma_l$ the ensemble of sequences of length $l$ built up by a moving windows of length $l$ i.e., if $\eta \in \Sigma_l$ then $\eta = a_i a_{i+1} \cdots a_{i+l-1}$ for some value of $i$. We calculate the values of $C(l)$ using this kind of ensemble $\Sigma_l$. As we will show by our simulations $C(l)$ scale as:

$$C(l) = l^{\alpha(m,s,\varphi)} \tag{5}$$

where $m$ is the memory size of the agents, $s$ is the number of strategies each of them bear and $\varphi$ (to be defined later) is a measure of how sophisticated is the behavior of the agents. We borrow some definitions of [9]. Let $U_{s,i}(t)$ be the score that agent $i$ assigns to strategy $s$ in the iteration $t$ of the game. These scores are updated in the additive way:

$$U_{s,i}(t+1) = U_{s,i}(t) + \Delta_{s,i}^{\mu(t)}(t, s_i, s_{-i}) \qquad (6)$$

where the term $\Delta_{s,i}^{\mu(t)}(t, s_i, s_{-i})$ quantifies the perceived succes of the strategy $s$ if agent $i$ played $s_i$ and the "aggregate" choice of the other agents were $s_{-i}$. The value $\mu(t)$ represent the state of the system at time $t$.

In the Minority Game the agents are naive. They only know the payoff delivered by the strategy $s_i(t)$ which they actually played. Hence, naive agents neglect their impact on the macroscopical observable $A^{\mu}(t)$. To be precise:

$$\Delta_{s,i}^{\mu(t)}(t, s_i(t), s_{-i}(t)) = -\frac{a_{s,i}^{\mu(t)} A^{\mu(t)}(t)}{2^m} \qquad (7)$$

where $a_{s,i}^{\mu(t)}$ is what would have played agent $i$ if she would have selected strategy $s$ and the state of the system would have been the real one, labeled here by $\mu(t)$.

In [9] the following correction to Eq. (7) is proposed:

$$\Delta_{s,i}^{\mu(t)}(t, s_i(t), s_{-i}(t)) = -\frac{a_{s,i}^{\mu(t)} A^{\mu(t)}(t)}{2^m} + \varphi \frac{\delta_{s,s_i(t)}}{2^m} \qquad (8)$$

where the added term models the tendency of the agents to stick to the strategy they are currently using. If $\varphi = 0$ then we have Eq. (7).

Notice that agents using Eq. (7) behave as "price takers" [9]. They do not take into account their influence in the macroscopical observable $A^{\mu}$. Agents using Eq. (8) repair in some sense this deficiency. Hence, a question naturally arise: Why don't reward to all strategies which give as outcome the same result as that actually played? More than that, why don't reward to all strategies with a quantity proportional to the payoff the agent would have obtained if she would have played it? In fact, several of the strategies of one agent could give the same answer as that actually played. Beside, the only way in which agents "impact" the market in these models is by their contribution to macroscopical observable $A^{\mu}$. No rumors, no news spread among them. Therefore, we propose a more realistic way of reward strategies:

$$U_{s,i}(t+1) = U_{s,i}(t) - \frac{a_{s,i}^{\mu(t)}\left(A^{\mu(t)}(t) - a_{s_i(t),i}^{\mu(t)} + a_{s,i}^{\mu(t)}\right)}{2^m} \quad (9)$$

That is, reward every strategy with what it would have won (lost) if the agent would have played it. Agents following Eq. (9) are able of real complex counter-factual thinking and we called *completely sophisticated* agents. As stated in [9] concept sophisticated as oposed to naive is borrow from [15].

Notice that Eq. (9) can be written as:

$$U_{s,i}(t+1) = U_{s,i}(t) - \frac{a_{s,i}^{\mu(t)} A^{\mu(t)}(t)}{2^m} + \frac{a_{s,i}^{\mu(t)} a_{s_i(t),i}^{\mu(t)} - 1}{2^m} \quad (10)$$

Therefore, the correction to naive behavior to become completely sophisticated is the term $\frac{a_{s,i}^{\mu(t)} a_{s_i(t),i}^{\mu(t)} - 1}{2^m}$. We finally propose the following procedure to reward strategies:

$$U_{s,i}(t+1) = U_{s,i}(t) - \frac{a_{s,i}^{\mu(t)} A^{\mu(t)}(t)}{2^m} + \varphi \frac{a_{s,i}^{\mu(t)} a_{s_i(t),i}^{\mu(t)} - 1}{2^m} \quad (11)$$

If $\varphi = 0$ the agents rewarding procedure is naive. If $\varphi = 1$ the agents rewarding procedure is completely sophisticated. When $\varphi$ moves from 0 to 1 Eq. (11) goes through all rewarding procedures from naive to completely sophisticated.

We make extensive computer simulations for several values of $m, s$ and $\varphi$. For every $m = 2,\ldots,5$, $s = 2,\ldots,5$ and $\varphi \in \{0, 0.1,\ldots,1\}$ we repeat ten times the computer experiment and for each of them construct the correspongding physical complexity function $C(l)$. They fit very well to a power function for every values of $m, s$ and $\varphi$. The exponent $\alpha$ is always positive. In Fig. 1 can be seen the plot of $C(l)$ v.s. $l$ for $m = 4, s = 3, \varphi = 0.4$:

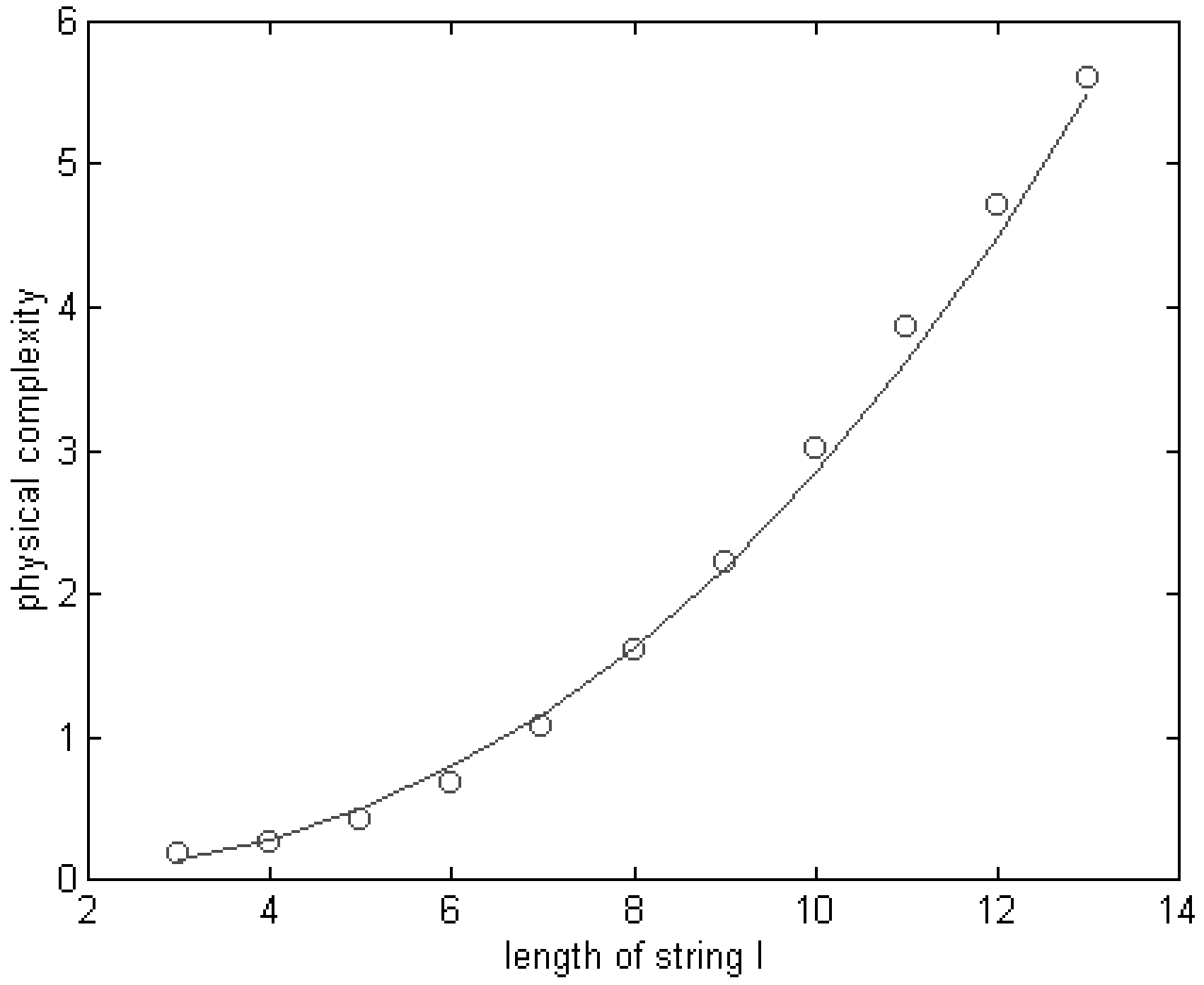

Fig. 1: Values of physical complexity v.s. the length of string of set $\Sigma_l$. The continuos line represent the fitness to a power function.

Later we averaged the values of exponents for each triplet $(m, s, \varphi)$ conforming a 3D array with entries $m, s$ and $\varphi$. With these data we performed a regression of exponent $\alpha$ with respect to the variables $m, s$ and $\varphi$ of the form:

$$\alpha(m, s, \varphi) = a(m, s)\varphi + b(m.s) \quad (13)$$

where:

$$a(m, s) = (a_1 m^2 + a_2 m + a_3)s^2 + (a_4 m^2 + a_5 m + a_6)s + a_7 m^2 + a_8 m + a_9$$

$$b(m,s) = (b_1 m^2 + b_2 m + b_3)s^2 + (b_4 m^2 + b_5 m + b_6)s + b_7 m^2 + b_8 m + b_9$$

The choice of this *ansatz* is based in our observations about the behavior of $\alpha$ with respect to $m$ and $s$ for every fixed $\varphi$. As we show later it is fully justified. The values of coeficients appear in the following table:

**Table I**

**Regresion coeficients in Eq. (13)**

| | | |
|---|---|---|
| $a_1 = 0.24450$ | $a_2 = -2.07810$ | $a_3 = 4.86120$ |
| $a_4 = -1.4423$ | $a_5 = 1.581800$ | $a_6 = 3.74100$ |
| $a_7 = 1.86670$ | $a_8 = -15.2176$ | $a_9 = 34.5322$ |
| $b_1 = -0.1696$ | $b_2 = 1.5818$ | $b_3 = -3.7410$ |
| $b_4 = 2.1426$ | $b_5 = -19.4888$ | $b_6 = 44.5105$ |
| $b_7 = -1.9711$ | $b_8 = 17.1166$ | $b_9 = -34.1939$ |

In the Fig. 2 we show a representation of function $a(m,s)$. The plot has been done continuously to enhance the behavior of $a(m,s)$. Notice that for every pair $(m,s)$ the value of $a(m,s)$ is negative. In our opinion it is related with a kind of efficient market hypothesis for this models . The efficient market hypothesis [16] establish that the market is highly efficient in the determination of the most rational prices. All available information is instantly processed when it reachs the market and it is immediately reflected in a new values of prices of assets traded. In our models, as $\varphi$ increases the agents reward more suitably each of their strategies. The value $\varphi = 1$ represent the perfect scenario in which each agent not only knows what he won (lost), but also what he would have won (lost) if he would have used the remainder strategies. Hence, as $\varphi$ increase the string $\varepsilon$ become more unpredictable and function $C(l)$ more flat. As we showed in [10] random sequences have flat function $C(l)$.

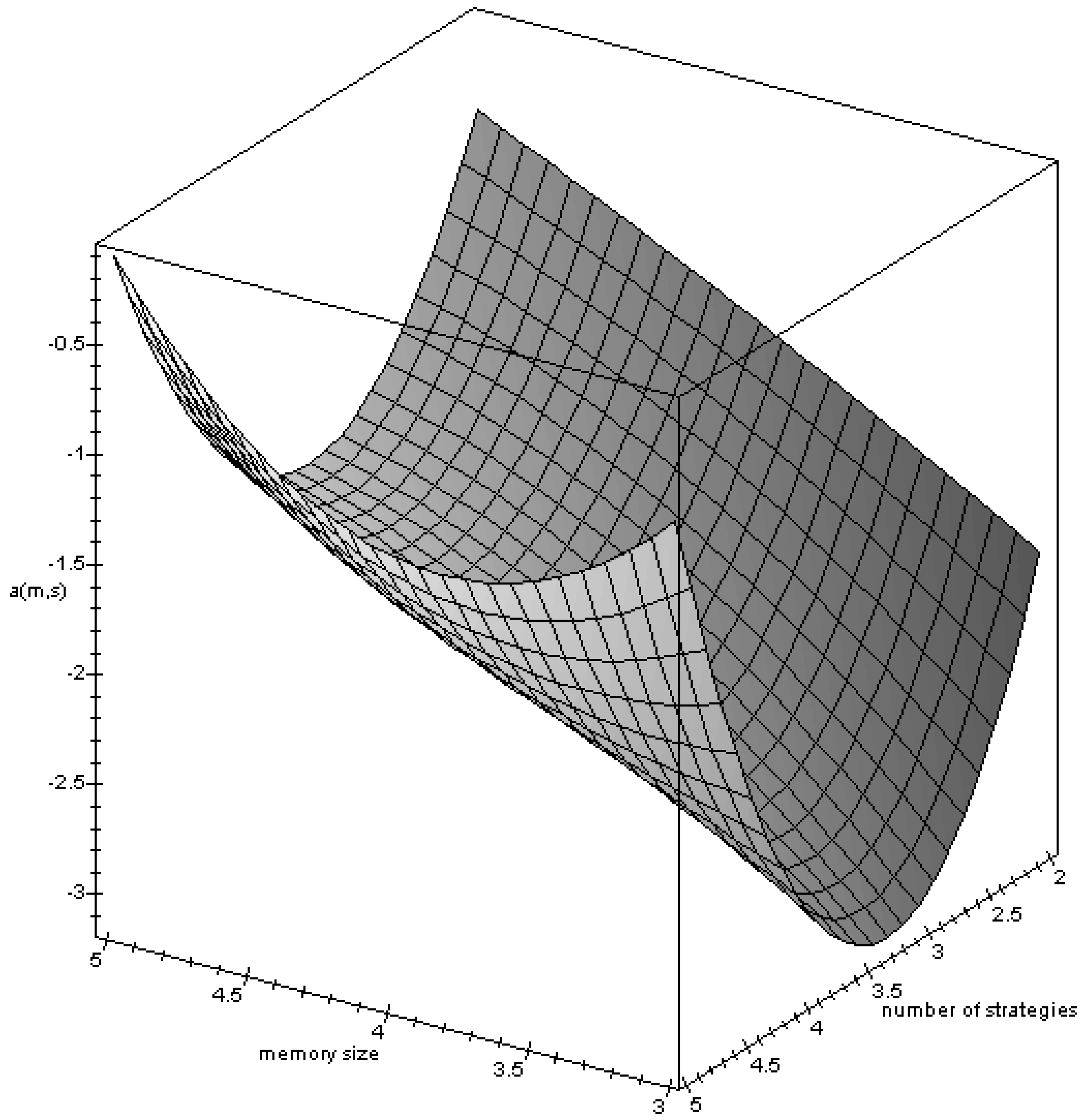

Fig. 2: The plot of the function $a(m,s)$. Notice that for every values of $m$ and $s$ the function is negative. The plot has been done continuously to enhance the behavior of the function.

Therefore, the function $a(m,s)$ can be interpreted as the rate of increase of randomness with respect to the increase of the "knowledge" of the agents $\varphi$. In the next section we study how changes in $\varphi$ affect the time correlation of macroscopical observable.

## 3 The behavior of mutual information function.

In the last section we show how the increase of $\varphi$ increases the randomness of the series of strings of the outcomes of the model . However, nothing has been said about the correlation of the outcomes along the time. Because the distance between two binary symbols in $\varepsilon$ represent the number of time iterations between them, a measure of the degree of correlation between elements in the symbolic string yield information about time correlation of the outcomes of winner room.

The mutual information function is defined as:

$$M(d) = \sum_{\alpha,\beta} P_{\alpha\beta}(d) \log_2 \left[ \frac{P_{\alpha\beta}(d)}{P_\alpha P_\beta} \right] \tag{6}$$

where: $P_{\alpha\beta}(d)$ is the probability of having a symbol $\alpha$ followed $d$ sites away by a symbol $\beta$ and $P_\alpha$ the density of the symbol $\alpha$. It can be proved (see [10] an references there in) that mutual information function is a very sensitive measure of correlation.

The most important feature of mutual information function of the string $\varepsilon$ is his remarkable persistence of high correlation at some large distances as well as sudden "falls" also at some large distance. For $\varphi$ small the lack of correlation accumulates yielding profound and periodic "falls" as can be seen in Fig. 3. In a market with these features the possibility of speculation would exist if some agents could anticipe these “falls”. We claim that this fact can be explained by the anomalous diffusion of the variable $\mu(t)$ [17] which is close related with the behavior of some cellular automata rules [18, 19]. We acknowledge the primacy of D.Chalet, M. Marsili and R. Zecchina [9] in the understanding of conection between behavior of $\mu(t)$ and Bruijn diagrams. As $\varphi$ increase two process develop due to the improve on the information level of the agents, as can be seen in Fig. 4: the profound "falls" become more sparsed. Speculation with great profit is less possible. Small "falls" appear more often. All of them are more smart and the margin of profit is smaller. When $\varphi \to 1$ almost no great "falls" appear and the series is very noisy.

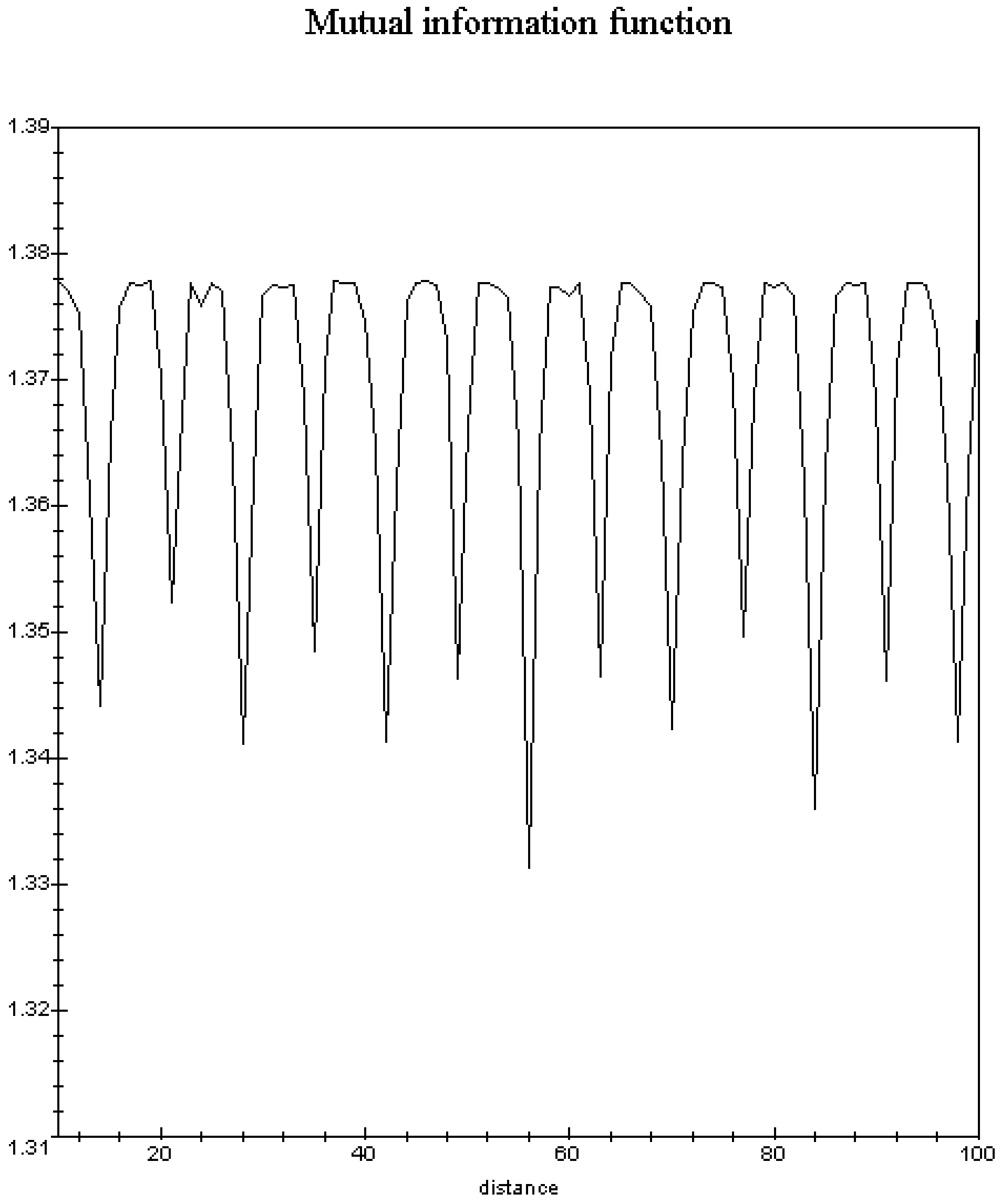

Fig. 3: Mutual information function for the values $m = 3, s = 2, \varphi = 0$. Notice the periodic behavior of the function.

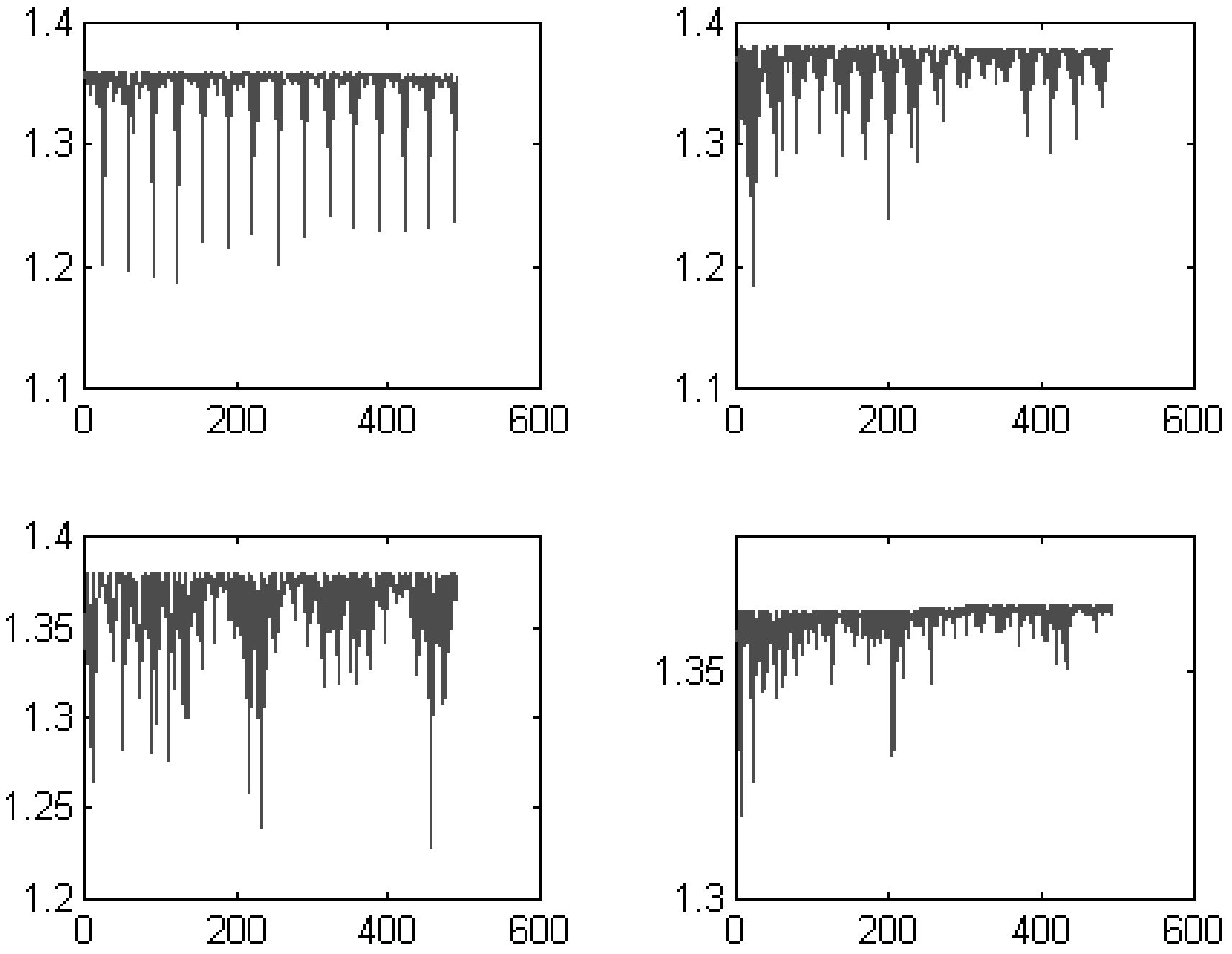

Fig. 4: The mutual information function for several values of $\varphi$ and $m, s$ fixed. Form top left to bottom right: $\varphi = 0.2$, $\varphi = 0.4$, $\varphi = 0.6$, $\varphi = 0.8$ . The other values are: $m = 4$, $s = 3$

## 4 Conclusions.

We have developed a more rational way of reward the different strategies of the agents. The parameter $\varphi$ is a measure of how good the agents understand the game and hence reward their strategies. We have shown that as $\varphi$ increase the number of binary digits which are predictable in reference to the whole series $\varepsilon$ decreases for every length $l$ of the string considered. It implies that the whole series $\varepsilon$ become more unpredictable as we showed in [10]. This behavior is reflected in the mutual information function of $\varepsilon$. The great "falls" appear less often, but small oscilations become more frecuently. It is close related with the increase of coordination of the agents. The *ansatz* proposed allow to obtain a measure of the loss of predictability as $\varphi$ increase: the coefficient $a(m,s)$. As we have shown, it is always negative. This can be interpreted as a kind of efficient market hypothesis because as far as the agents better incorporate the information about the outcomes the curve of $C(l)$ become more flat in agreement with a more random behavior.

## References.